\documentclass[aps,preprint,showpacs,superscriptaddress,prb]{revtex4}
\usepackage{times,xspace}
\usepackage{amsbsy,amssymb,amsmath,bm}
\usepackage{graphicx,color,epsfig,rotate}
\usepackage{fancyheadings}

\begin{document}
\title{VISCOELASTIC PROPERTIES OF DYNAMICALLY ASYMMETRIC BINARY FLUIDS UNDER SHEAR FLOW}
\author{Vinay Dwivedi}
\affiliation{Department of Mechanical and Environmental
Engineering, University of California, Santa Barbara California 93106}
\affiliation{Theoretical Division, Los Alamos National Laboratory,
Los Alamos, New Mexico, 87545}
\author{Rajeev Ahluwalia}
\author{Turab Lookman}
\author{Avadh Saxena}
\affiliation{Theoretical Division, Los Alamos National Laboratory,
Los Alamos, New Mexico, 87545 USA}
\date{\today}
\begin{abstract}
We study theoretically the viscoelastic properties of sheared
binary fluids that have strong dynamical asymmetry between the two
components. The dynamical asymmetry arises due to asymmetry
between the viscoelastic stresses, particularly the bulk stress.
Our calculations are based on the two-fluid
model that incorporates the asymmetric stress distribution. We
simulate the phase separation process under an externally imposed
shear and compare the asymmetric case with the usual phase
separation under a shear flow without viscoelastic effects. We
also simulate the behavior of phase separated stable morphologies
under applied shear and compute the stress relaxation.
\end{abstract}

\maketitle

\section{INTRODUCTION}
Many fluids can exhibit viscoelastic behavior, i.e. their response
to deformation is intermediate between that of solids and fluids.
For short times, the response is elastic and the stress is
proportional to the applied strain. On the other hand, in the long time limit,
a fluid like response with stress 
proportional to the strain rate is observed. The effect of 
viscoelasticity on the morphologies of phase separating
polymer solutions and blends has been recently experimentally
studied by Tanaka \cite{soltanaka,blendtanaka}. It is now well
established that due to an asymmetry between the viscoelastic
properties of the two phases, a transient network of the more
viscoelastic phase is observed. While there is some understanding
of the influence of viscoelastic effects on phase separation, little  
 is known on
how viscoelasticity effects
phase separation when an external shear flow is applied. Although
effects of shear on newtonian (purely hydrodynamic) binary fluids
have been well studied \cite{shexp,lookman} and dynamical steady
states with stringy morphologies have been observed, it is also
important to study similar situations for binary fluids where
viscoelastic effects can dominate.

Theoretically, the coupling between the stress and concentration
fluctuations was investigated by Helfand and Fredrickson
\cite {fred} in the context of shear flow in polymer
solutions. Doi and Onuki \cite{doi} studied this coupling by
introducing a two-fluid model that considers two different
velocities for the two components. Taniguchi and Onuki
\cite{Onuki} have used the two-fluid model to study viscoelastic
phase separation. However, a two-fluid model incorporating
dynamical asymmetry due to asymmetric distribution of bulk
stresses 
\cite{araki2d,araki3d,revtanaka,china1} 
has successfully explained
many
features of the viscoelastic phase separation experiments of
Tanaka, such as the transient network formation and eventual phase
inversion. Recently, this theory has also been applied to study
the viscoelastic phase separation in diblock copolymers
\cite{china2}. The two-fluid model was also used to simulate
polymeric foams under shear \cite{grant}. In this paper, we use
this model to study phase separation under a shear flow for a
binary fluid that is characterized by a strong asymmetry between
the viscoelastic moduli of the two phases, particularly the bulk
modulus.

In addition to phase separation under shear flow, the behavior of
stable phase separated states under a deformation is also
important. Recently, Buxton and Balazas \cite{buxton} have studied
the mechanical properties of phase separated morphologies using a
block and spring model. This model can only model elastic
properties and cannot describe the viscoelastic response. Thus we
study the behavior of fully phase separated stable states under a
constant shear based on the two-fluid model. The time evolution of
mechanical stresses generated for such phase separated states is
also studied.

The paper is organized as follows. In Sec. 2, we discuss the
two-fluid model used in our simulations. Section 3 describes
results on viscoelastic phase separation under shear flow while
section 4 is devoted to the response of stable phase separated
morphologies subjected to steady shear. We conclude the paper 
with a summary and discussion in Sec. 4.
\section{TWO-FLUID MODEL}
The coupling between viscoelastic stresses and diffusion is
studied using the two-fluid model. We consider a dynamically
asymmetric mixture of a viscoelastic fluid $(A)$ and a purely
hydrodynamic (non-viscoelastic) fluid $(B)$. In this model,
different velocity fields, $\vec{v_A}(\vec{r},t)$ for the viscoelastic
phase and $\vec{v_B}(\vec{r},t)$ for the non-viscoelastic fluid are
introduced. The average velocity is given by
$\vec{v}_{average}(\vec{r},t)={\phi(\vec{r},t)}\vec{v_A}(\vec{r},t)
+({1-\phi(\vec{r},t)})\vec{v_B}(\vec{r},t)$, where $\phi(\vec{r},t)$
is the concentration of the viscoelastic fluid. The free energy of
mixing can be written in terms of the concentration
$\phi(\vec{r},t)$ as
\begin{equation}
F_{mix}=\int d\vec{r}[{\phi}^2{(1-\phi)}^2 +(C/2)({\vec{\nabla}}\phi)^2],
\end{equation}
where $C$ is the concentration gradient coefficient that is assumed to be
constant in the present work. The free energy $F_{mix}$ is a
simplified version of the Flory-Huggins \cite{flory} type free
energy.  The applied shear flow is implemented by considering an average 
velocity 
$\vec{v}_{average}=\vec{v}_{applied}+\vec{v}$, where $ \vec{v}$ is the contribution to 
the average velocity 
due to 
concentration fluctuations and $\vec{v}_{applied}$ is the contribution due to 
the external 
shear flow given as, 
\begin{equation}
\vec{v}_{applied}=s|x-x_{0}|\hat{j}.
\end{equation}
This represents an external shear (symmetric about $x_{0}$) 
applied along the  $y$ direction and $s$ is the shear rate. 
The appropriate equations of motion for incompressible
viscoelastic binary fluids under external shear flow are 
\begin{equation}
{{\partial \phi}\over{\partial t}}+\vec{v}\cdot{\vec{\nabla}{\phi}}
+s|x-{x_{0}}|
{{\partial \phi}\over{\partial y}}
={\vec{\nabla}}\cdot{ {{\phi}(1-\phi)^2}\over{\zeta} }
[\vec{\nabla}\cdot{\stackrel{\leftrightarrow}\Pi}
-\vec{\nabla}\cdot{\stackrel{\leftrightarrow}\sigma}].
\end{equation}
Here $\stackrel{\leftrightarrow}\sigma$ represents the
viscoelastic stress tensor.
The osmotic tensor
$\stackrel{\leftrightarrow}\Pi$ is defined as 
$\vec{\nabla}\cdot
\stackrel{\leftrightarrow}\Pi=
\phi\vec{\nabla}\{2{\phi}(1-\phi)^2-2(1-\phi){\phi}^2
-C{\nabla}^2{\phi}\}$. The quantity $\zeta$ is the friction
coefficient associated with the velocity $\vec{v_A}-\vec{v_B}$.
The dynamics of the average velocity field $\vec{v}$ is given by the usual
Navier-Stokes equation
\begin{equation}
\rho
{{\partial \vec{v}}\over{\partial t}}=
-\vec{\nabla}\cdot \stackrel{\leftrightarrow}\Pi+\vec{\nabla} P+
\vec{\nabla}\cdot\stackrel{\leftrightarrow}\sigma +
{\eta_s}{\nabla}^2{\vec{v}},
\end{equation}
where a pressure has been introduced to account for the
incompressibility constraint. If we use the overdamped limit
${{\partial \vec{v}}\over{\partial t}}=0$, then under the
incompressibility condition $\vec{\nabla}\cdot{\vec{v}}=0$, the
velocity can be expressed in Fourier space as
\begin{equation}
\vec{v_k}={\stackrel{\leftrightarrow}T_k}\cdot(-\vec{\nabla}\cdot\stackrel{\leftrightarrow}\Pi+
\vec{\nabla}\cdot\stackrel{\leftrightarrow}\sigma)_{k},
\end{equation}
where $()_k$ represents the Fourier transform and $T_{k}$ is a tensor defined in
fourier space as
\begin{equation}
{\stackrel{\leftrightarrow}T_k}={{1}\over{{\eta_s}{k^2}}}
\bigg[ \stackrel{\leftrightarrow}I-{{\vec{k}\vec{k}}\over{k^2}}\bigg],
\end{equation}
where $\stackrel{\leftrightarrow}I$ represents the unit tensor.
Next, we have to specify a constitutive law for the viscoelastic
stresses. We use the Maxwell model according to which the equation
of motion for the stresses is given by 
\begin{eqnarray}
{{\partial \stackrel{\leftrightarrow}\sigma_b}\over{\partial t}}
+{{\stackrel{\leftrightarrow}\sigma_b}\over{{\tau_b} (\phi)}} &=&
m_b(\phi)\stackrel{\leftrightarrow}M,
\nonumber \\
 {{\partial
\stackrel{\leftrightarrow}\sigma_s}\over{\partial t}}
+{{\stackrel{\leftrightarrow}\sigma_s}\over{{\tau_s} (\phi)}} &=&
m_s(\phi)\stackrel{\leftrightarrow}M,
\end{eqnarray}
where $\tau_b$ and $\tau_s$ are the internal molecular relaxation
times associated with the bulk and shear stresses respectively.
The matrix $\stackrel{\leftrightarrow}M$ is given as
\begin{equation}
\stackrel{\leftrightarrow}M=\left[ \begin{array}{cc} 2{{\partial
{v_A}_{x}}\over{\partial x}} & ({{\partial
{v_A}_{x}}\over{\partial y}}+ {{\partial
{v_A}_{y}}\over{\partial x}})\\
({{\partial {v_A}_{x}}\over{\partial y}}+ {{\partial
{v_A}_{y}}\over{\partial x}})& 2{{\partial
{v_A}_{y}}\over{\partial y}}\end{array} \right].
\end{equation}
The velocity $\vec{v_A}$ represents the velocity of the
viscoelastic phase defined as
\begin{equation}
\vec{v_A}=\vec{v}+\vec{v}_{applied}-
{{(1-\phi)^2}\over{\zeta}}\left[\vec{\nabla}
\cdot\stackrel{\leftrightarrow}{\Pi}
-\vec{\nabla}\cdot\stackrel{\leftrightarrow}\sigma\right].
\end{equation}
The above definition of $\stackrel{\leftrightarrow}M$ assumes that
the stress acts on the viscoelastic fluid only. 
The final stresses $\stackrel{\leftrightarrow}{\sigma_b}^{f}$ and
$\stackrel{\leftrightarrow}{\sigma_s}^{f}$ are calculated as
$\stackrel{\leftrightarrow}{\sigma_b}^{f}=
(1/2)Tr(\stackrel{\leftrightarrow}{\sigma_b})
{\stackrel{\leftrightarrow}{I}}$
and 
$\stackrel{\leftrightarrow}{\sigma_s}^{f}=
\stackrel{\leftrightarrow}{\sigma_s}-
(1/2)Tr(\stackrel{\leftrightarrow}{\sigma_s})\stackrel{\leftrightarrow}{I}$.
The total stress is calculated as
$\stackrel{\leftrightarrow}{\sigma}=
\stackrel{\leftrightarrow}{\sigma_b}^{f}+
\stackrel{\leftrightarrow}{\sigma_s}^{f}$. The model decribed in
equations $(1)$ to $(9)$ can be used to simulate the viscoelastic phase
separation under shear flow with shear rate $s$.
\section{VISCOELASTIC PHASE SEPARATION UNDER SHEAR FLOW}
We describe our results on simulations of
viscoelastic phase separation, both with and without shear. The
model described in Sec. II is discretized on a
$128 \times 128$ grid with periodic boundary conditions, using a 
spectral approach for the velocity equation and explicit, central 
difference schemes $(\Delta x=1,\Delta t= 0.005)$ for the rest. 

Following Tanaka and Araki \cite{araki2d,araki3d}, we choose
$m_b={m_b}^{0}\theta(\phi-{\phi_0})$ ($\phi_0$ is the initial
concentration before quenching), $m_s={m_s}^0{\phi}^2$,
${\tau_b}={\tau_b}^0{\phi}^2$ and ${\tau_s}={\tau_s}^0{\phi}^2$.
With this choice of the moduli and time scales, Tanaka and Araki
were able to reproduce  a number of experimental features of viscoelastic
phase separation \cite{araki2d,araki3d}.

We first simulate phase separation without an external
shear$(s=0)$. We consider a $50-50$ mix 
with initial concentration fixed at  $\phi_0=0.5$. The field
$\phi(\vec{r},0)$ is initialized by small fluctuations of order 0.001 around
$\phi_0$ and the model in Sec. II is simulated for three
different cases. Case I corresponds to hydrodynamic phase
separation in the absence of viscoelastic effects. Here we set all the
stresses to zero and solve Eqs. (3) and (5) with only one
velocity field $\vec{v}$. Case II is the full asymmetric
viscoelastic model with $\eta_s=1$, $\zeta=0.1$, ${m_b}^0=5$, ${m_s}^0=0.5$, 
${\tau_b}^{0}=10$ and ${\tau_s}^{0}=50$. To clarify the role of the bulk stress, in case III we
switch off the bulk modulus, i.e., ${m_b}^{0}=0$.

Figure 1 shows the phase separation process for the three cases.
The left column shows the non-viscoelastic (hydrodynamic) case,
middle column depicts the full viscoelastic model with asymmetric
bulk modulus and the right column represents the phase separation
for viscoelastic case without bulk modulus. This figure shows the
importance of the bulk modulus in the formation of a network of
the more viscoelastic component. For the case with the bulk modulus,
initially, holes of the non viscoelastic phase are formed in a
partially phase separated matrix of the more viscoelastic phase.
As these holes grow, the area of the viscoelastic region
decreases and a network like structure is formed. These results
are consistent with earlier results of Tanaka and Araki
\cite{araki2d}. 
For the full viscoelastic model, we do not
observe the breaking of the network within the time interval
we simulated. The case without the bulk modulus appears to be
similar to the purely hydrodynamic case and there is no network
formation. Figure 1 also shows that viscoelastic effects slow down
the domain growth, as can be inferred by comparing the domain
patterns at time $t=1000$ for all the three cases.

We have also monitored the evolution of the viscoelastic stresses
during the phase separation. Figure 2 shows the behavior of the average
normal stress $\sigma_{xx}-\sigma_{yy}$, the average bulk stress
$\sigma_{xx}+\sigma_{yy}$ and the average shear stress $\sigma_{xy}$ for
the case with bulk stress as well as that for the case without bulk
stresses. The normal and shear stresses do not decay completely in
the time simulated and  exhibit small fluctuations of order $0.001$, possibly due to the
motion of the viscoelastic phase. The behavior of the bulk stress
is worth noticing as for the case with a bulk modulus, the early
stage of phase separation is associated with significant
compressive stresses. The compressive stresses are
associated with the formation of the network of the more
viscoelastic phase. The network of partially phase separated
viscoelastic phase shrinks as the phase separation proceeds
resulting in the compressive stresses observed in the early
stages.

In Fig. 3, we display the phase separation process when an
external shear with shear rate $s=0.01$ is applied. As in the case
without shear, we consider three cases corresponding to the
hydrodynamic case without viscoelastic effects, full viscoelastic
model with bulk modulus and viscoelastic model without bulk
modulus. At this shear rate, the very early stages are not
influenced significantly by the applied shear for all the cases.
At later times, there is an underlying tendency for the
domains to align along the shear direction, although the alignment
is not complete within the time simulated $(t=2500)$. The case
with the bulk modulus exhibits network morphologies in the earlier
stages even for this case. The alignment is also slower in the
initial stages for the case with the bulk modulus.

The time evolution of viscoelastic stresses for this value of
shear is shown in figure 4 for the case with bulk stress as well as
case without bulk stress. Fluctuations due to random motion of
domains in the early stages are observed for both normal and shear
stresses and the amplitude of the fluctuations decreases at late
times as the domains tend to align along the flow direction. For
this case also, the bulk stress for the case with bulk modulus becomes
negative in the early stages, indicating the shrinking of the
network of the viscoelastic phase.

Finally, we consider phase separation under a relatively large
shear rate $s=0.1$. Figure 5 shows the phase separation at this
shear rate for all three cases. For this case, the external shear
effects dominate the phase separation and the tendency for the
domains to align along the shear to form stringy patterns is much
stronger than the earlier case. The purely hydrodynamic model
rapidly forms the stringy phase, whereas for the viscoelastic
cases, the intermediate stages are characterized by complex motion of
the domains.  For the case with the bulk modulus, the early stages
do not show the formation of a well developed network as the shear
dominates over the internal dynamics. The snapshots at $t=2500$ show
that a stringy phase seems to form at long times for the
viscoelastic cases also, although the width of the lamellae is
larger compared to the hydrodynamic case.

Figure 6 shows the time evolution of the viscoelastic stresses for
this case. After transient fluctuations, the normal stresses decay
to zero as the steady state is established. Compressive bulk
stresses are observed even for this case but the magnitude is 
smaller for this case, compared to the earlier cases. This is due
to the fact that the shear flow is so fast that the transient
network of the viscoelastic phase is not allowed to fully develop.
This is clear from the snapshot at time $t=50$ for the
viscoelastic case. The behavior of the shear stresses is
interesting. At long times, the shear stresses
saturate to non zero values. The saturation stresses are related
to the effective shear viscosities for this binary fluid.
\section{PHASE SEPARATED STABLE STATES UNDER SHEAR}
So far we have considered the effects of shear on binary
fluids undergoing phase separation. In this section we apply
shear on stable phase separated morphologies. The aim of this
section is to explore the possibilities of using this model to
study mechanical properties of polymer solutions and blends.
Recently, Buxton and Balazs \cite{buxton} have used a block and
spring model to simulate deformation of a random two-phase
morphology obtained from a Cahn-Hilliard simulation. In this
approach, there is no coupling between the morphologies and the
deformation. Such an approach can only be used to study linear
elastic behavior and does not describe the full viscoelastic
response. The morphologies cannot evolve in response to
deformation. However, the two-fluid framework used in this paper
explicitly incorporates a coupling between the stress and the
concentration and can describe deformation induced morphological
evolution.

Here, we consider a simple phase separated morphology
below the coexistence temperature. We consider a $50-50$
$(\phi_0=0.5)$ binary fluid that exists as a band of the fluid
$A$ sandwiched between layers of fluid $B$. This is a stable
phase separated configuration. Polymer blends can show such 
macro-phase separation in equilibrium and block copolymers also 
 micro-phase separate into lamellar morphologies \cite{hamley}.
We apply shear in the same way as
discussed in the previous section. We consider two different
shearing conditions. In one case the shear is normal to the $A/B$
interface and in the other case, the shear is parallel to the
interface.  Such a situation (for block copolymers) has been investigated  
experimentally 
\cite{nature} and theoretically using hydrodynamic models \cite{drolet}. 
The morphological evolution for the case of shear
normal to the $A/B$ interface is depicted in Fig. 7. All three
cases that have been discussed in the previous section are simulated for this case
also. For all the cases, initially the interface starts to move in
response to the shear and at long times there is a tendency to
form many bands aligned along the shearing direction. The bands
appear to be wider for the viscoelastic cases in comparison to the
hydrodynamic case. However, we do not find much difference between
the case with bulk modulus and the case without bulk modulus.
Interestingly, for the case when the shear is applied parallel to
the interface, the band remains stable and no morphological
evolution is observed for all the three cases. This is consistent with the 
shear induced reorientation of lamella for block copolymers observed in experiments \cite{nature} 
and theoretical models \cite{drolet}.

Figure 8 shows the evolution of stresses for the full viscoelastic
model (with bulk modulus) for both cases (shear parallel and
perpendicular to the interface). For shear parallel to the
interface, the normal and bulk stresses remain zero during the
shearing process. This is due to the fact that there are no
morphological changes and only shear stresses are generated. For
the case of shear normal to the interface,  normal as well as bulk
stresses are generated due to shear induced motion of the
interface depicted in Fig. 7. For a short interval, weak
compressive stresses are observed. The early time behavior of
the shear stress also differs from the case when the shear
is applied parallel to the interface. This figure demonstrates the
role played by the morphological evolution on the viscoelastic
properties of binary fluids. For the case without bulk modulus, very 
similar stress evolution (not shown here) is observed. However, no bulk 
stress is generated.

Finally, we study the relaxation of the morphologies after the
applied shear is suddenly removed. We remove the shear at $t=50$
corresponding to the morphologies shown in the bottom row of
Fig. 7. Fig. 9 shows the final morphologies at $t=2150$ after
removing the shear. It is clear that the morphological relaxation
is much faster for the non-viscoelastic case and the band
reappears, although it is shifted from the original position. For
the viscoelastic cases, the morphological relaxation is very slow
and the pattern does not evolve to the original band within the
time span of the present simulations.
\section{SUMMARY}
We have investigated the viscoelastic properties of
dynamically asymmetric binary fluids under shear flow. The
dynamical asymmetry arises as the binary fluid constitutes one
viscoelastic fluid and the other non-viscoelastic purely
hydrodynamic fluid. We have simulated phase separation for this
binary fluid both with and without shear. Only for the case with a
nonzero bulk modulus, there is a tendency to form a network of
the more viscoelastic phase in the initial stages of phase
separation, both with and without shear flow. However, for the
high shear rate, the network phase is short lived as the shear has
a tendency to suppress network formation. For the high shear rate,
in the long time limit, a stringy phase is observed even for the
viscoelastic cases, however the length scales associated with the
stringy phase are larger compared to the purely hydrodynamic case.
Thus, viscoelastic effects enhance the extent of phase separation
under shear. We have also studied the temporal evolution of the
effective viscoelastic stresses during the phase separation
process. Interfacial motion results in fluctuations of the shear
and normal stresses in the early stages. Transient compressive
stresses are also observed corresponding to shrinking of
viscoelastic phase in the early stages. At long times, the shear
stresses saturate to a nonzero value. This value is related to
the effective viscosity of the two-phase fluid.

We have investigated the effects of shear on stable phase
separated morphologies. A lamellar structure of the binary fluid
below the coexistence temperature is sheared, both normal and
parallel to the interface. There are crucial differences between
these two cases. For the shear parallel to the interface, no
morphological evolution is observed while for the case of shear
normal to the interface, the interface moves and splits into bands
that tend to align along the shear. The evolution of effective
stresses also depends on the direction of the interface. For
example, for the shear parallel to the interface, no  bulk and 
normal stresses are generated. On the other hand, for the shear
normal to the interface, bulk, normal and shear stresses are
generated due to interfacial motion.

This paper demonstrates the role played by viscoelasticity on
shearing of binary fluids. The simulations show that the
underlying viscoelastic properties can significantly influence the
shearing behavior. Further experiments are needed to test
predictions of our simulations. The approach used in the
present work can also be applied to other complex fluids such as
block copolymers and microemulsions.
\begin{acknowledgments}
V.D. is grateful to the LANL-UCSB CARE program for providing 
support through a summer student stipend. This work was 
supported by the US Department of energy.
\end{acknowledgments}

\newpage
Figure Captions:

1. Phase separation process without shear flow for the purely 
hydrodynamic case, 
viscoelastic case with bulk modulus and viscoelastic case without 
bulk modulus.
  
2. Evolution of the normal, bulk and shear stresses for the 
situation depicted in Fig. 1. The stresses are calculated by averaging 
the local stresses over the entire system. 

3. Phase separation process with shear flow of shear rate 
$s=0.01$ for the purely 
hydrodynamic case, 
viscoelastic case with bulk modulus and viscoelastic case without 
bulk modulus. 
Shear is applied along the $y$ direction and periodic 
boundary conditions are assumed.
  
4. Evolution of the normal, bulk and shear stresses for the 
situation depicted in Fig. 3.

5. Phase separation process with shear flow of shear rate 
$s=0.1$ for the purely 
hydrodynamic case, 
viscoelastic case with bulk modulus and viscoelastic case without 
bulk modulus.  
Shear is applied along the $y$ direction and periodic 
boundary conditions are assumed.

6. Evolution of the normal, bulk and shear stresses for the 
situation depicted in Fig. 5.

7. Evolution of the band morphology under a shear flow of shear rate 
$s=0.1$ for the purely 
hydrodynamic case, 
viscoelastic case with bulk modulus and viscoelastic case without 
bulk modulus.  
Shear is applied along the $y$ direction and periodic 
boundary conditions are assumed.  

8. Evolution of the normal, bulk and shear stresses for the 
 case with nonzero bulk modulus (evolution depicted in the middle column of Fig. 7).

9.  Morphologies for the three cases at $t=2150$ time steps after the
 shear flow has been removed. The shear flow was removed at $t=50$ steps after  the shear was applied (corresponding to the last row in Fig. 7). 
\end{document}